# Classifying Application Phases in Asymmetric Chip Multiprocessors

A. Z. Jooya

Computer Science dept.
Iran University of Science and Technology
Tehran, Iran
.

M. Analoui

Computer Science dept.
Iran University of Science and Technology
Tehran, Iran
.

*Abstract—* **In present study, in order to improve the performance and reduce the amount of power which is dissipated in heterogeneous multicore processors, the ability of detecting the program execution phases is investigated. The program's execution intervals have been classified in different phases based on their throughput and the utilization of the cores. The results of implementing the phase detection technique are investigated on a single core processor and also on a multi-core processor. To minimize the profiling overhead, an algorithm for the dynamic adjustment of the profiling intervals is presented. It is based on the behavior of the program and reduces the profiling overhead more than three fold. The results are obtained from executing multiprocessor benchmarks on a given processor. In order to show the program phases clearly, throughput and utilization of execution intervals are presented on a scatter plot. The results are presented for both fixed and variable intervals.**

## I. INTRODUCTION

The amount of diversity among applications that a typical computer is expected to run can be considerable. Also there is significant diversity among different phases of the same application. The utilization of processor resources changes among different phases (a program phase is defined as a contiguous interval of program execution in which the program behavior remains relatively unchanged). Therefore, if a processor had the ability of detecting phase changes, it could adapt its resources based on the new phase requirement. This can cause significant reduction in power consumption [1].

In [2] authors proposed positional adaptation which uses the program structure to identify major program phases. They proposed the use of a hardware based call stack to identify program subroutines.

In [3] authors found a relationship between phases and instruction working sets, and they found that phase changes occur when the working set changes. In [4, 5], authors used techniques from machine learning to classify the execution of the program into phases (clusters).

In [6] authors improved the execution time and power of multicore processors by predicting the optimal number of threads depending on the amount of data synchronization and the minimum number of threads required to saturate the off-chip bus.

In [7] accelerated critical sections (ACS) technique is introduced which leverages the high-performance core(s) of an Asymmetric Chip Multiprocessor (ACMP) to accelerate the execution of critical sections.

In [8] authors proposed scheduling algorithms based on the Hungarian Algorithm and artificial intelligence (AI) search techniques which effectively match the capabilities of each core with the requirements of the applications.

In [9] authors found that the use of ready and in-flight instruction metrics permits effective co-scheduling of compatible phases among the four contexts.

In [10] authors proposed a scheme for assigning applications to appropriate cores based on the information which the job itself presents as an architectural signature of the application, and is composed of certain microarchitecture-independent characteristics.

In [11] authors made a case that thread schedulers for heterogeneous multicore systems should balance between three objectives: optimal performance, fair CPU sharing, and balanced core assignment. They argued that thread to core assignment may conflict with the enforcement of fair CPU sharing. In [12] they introduced a cache-fair algorithm which ensures hat the application runs as quickly as it would under fair cache allocation, regardless of how the cache is actually allocated. Comparison between our proposed scheme and previous studies is part of our ongoing research.

In heterogeneous multi-core processors, when the phase change is detected, the process will be switched to the core for which the performance is much closer to the new phase requirement. This strategy could best prevent the core resource under utilization or over utilization. In the present study, the phase changes of the program's processes that are distributed between cores are investigated, focusing on multiprocessor benchmarks' phase behavior in multicore processors.

The rest of the paper has the following organization. In section two, the methodology is described for fix and variable profiling interval length. How the throughput and utilization of intervals are used to detect the phase changes and adjusting the interval lengths dynamically are described. Section three presents simulations results for single and multicore processors and section four summarizes the results and contributions of our work.

## II. METHODOLOGY

When a process starts running on a core, it goes through different phases during its execution and each phase has specific demand on the core. Therefore, the core selected to run a process will not remain the best for all phases. This leads the scheduler to make assignment decision in a dynamic manner. The goal of such a scheduler is assigning each process to a core which best satisfy its requirements and prevents under-utilization or over-utilization of core's resources.

Here presented phase detection technique uses dynamic profiling with variable profiling interval length. The profiling starts with minimum interval length and increases based on program's behavior. This causes significant reduction in profiling overhead.

To detect a phase change, the throughput (the number of instructions retiring in an interval) average of all intervals which belong to the same phase is calculated and the difference percentage of current interval throughput from throughput average of previous intervals of the phase, $D_i$, is considered as phase detection criterion, and defined as follow:

$$D_i = \frac{(Th_i - \overline{Th_{i-1}}) \times 100}{\overline{Th_{i-1}}} \qquad (1)$$

in which $Th_i$ is the throughput of $i^{th}$ interval and $\overline{Th_i - 1}$ is The throughput average of previous intervals that belong to the same phase and, for a given i, defined as follow:

$$\overline{Th_i} = \frac{Th_i + (\overline{Th_{i-1}} \times N_{i-1})}{N_i} \qquad (2)$$

in which, $N_i$ is the number of phase intervals.

The phase change is detected by comparing $D_i$ with a predefined threshold, $\delta_{Th}$, for throughput changes. The utilization (of integer and floating point functional units) that is obtained for each interval are profiled separately. The higher utilization is chosen for each interval, because in the most of benchmarks, there are significant difference between the number of integer and floating point instructions, and usually one of them overcomes the other. Comparing the utilization of each interval, $U_i$, with predefined thresholds, $\delta_{over\_utilize}$ and $\delta_{under\_utilize}$, for detecting core resource over utilization or under utilization, respectively. Fig. 1 presents the relevant pseudo-code which summarizes the above discussion of using the throughput and utilization of intervals to make a decision about phase changes.

```
If ( U_i < δ_under_utilize )
        under_utilization_flag = 1;
else if ( U_i > δ_over_utilize )
        over_utilization_flag = 1;
if (|D_i| > δ_Th || under_utilization_flag || over_utilization_flag)
        intervals are not similar;
else
        intervals are similar;
```

Figure 1.   The pseudo-code for similarity function.

As previously described, the interval length, $\tau$, is better to be adjusted dynamically in order to reduce the profiling overhead. The related algorithm is illustrated in Fig. 2. In this study, the minimum of 100K cycles is assumed as the default value of $\tau$. A phase is called steady if the difference between throughput average of all previous intervals and throughput average of all intervals doesn't meet the predefined threshold. In this case the interval length is doubled. Otherwise, the interval length is divided by two. It is worth noting that all thresholds, used in the algorithm, have been obtained after numerous simulations.

```
Phase_chang_flag=0;
if ( |Th_i - Th_{i-1}| < throughput_average_chang_threshold )

        steady_phase_count++;
        if ( steady_phase_count == steady_phase_upper_bound )
              τ *= 2 ;
else

        steady_phase_count = 0;
        τ /= 2 ;
```

Figure 2.   The pseudo-code for adjusting interval length dynamically.

The utilization and throughput of program intervals are shown with scatters. This method of representation is more beneficial to distinguish different phases. The ideal points in the throughput-utilization scatter are those with both high utilization and throughput. Approaching the utilization to 100 % indicates that core resources are over-utilized. This condition implies that the application could use more resources and that switching to a stronger core will most likely boost performance. On the other hand, points with low utilization indicate that the core's resources are under-utilized. In this condition switching the application to a weaker core will most likely reduce power dissipation without compromising performance.

### A.  Microarchitecture

The simulation has been performed on four cores of two types with different level of performance. Table 1 summarizes the characteristics of the cores. The A-type cores have the higher performance. The level one cache is private for each core and the instruction and data caches are separated. A large unified level two cache is shared between all the cores. MOSE protocol is used for cache coherency. This architecture of memory hierarchy is the most commonly used one in multi-core processors. We consider that all the cores are implemented in 100 nm technology and run at 2.1 GHz.

TABLE I.          CHARACRISTICS OF THE CORES.

| Core | A-type | B-type |
|------|--------|--------|
| Issue-width | 4 ( Out-Of-Order ) | 2 ( Out-Of-Order ) |
| IL1-cache | 64 KB , 4 way | 32 KB , 2 way |
| DL1-cache | 64 KB , 4 way | 32 KB , 2 way |
| L2-cache | 4 Mb , 8 way | (shared) |
| B-predictor | hybrid | hybrid |
| Int window size | 80 | 56 |
| FP window size | 32 | 16 |

### B.  Simulator and Benchmarks

The simulations have been carried out utilizing SESC simulator developed by Jose Renau et al. [13]. This simulator can probably model a very wide set of architectures such as single processors, multicore processors and thread level speculation. Authors made the required modifications to include the phase detection algorithm.

Four scientific/technical parallel workloads from splash2 [14] have been used. These workloads consist of three applications and one computational kernel. The kernel is FFT and the three applications that we have used are Barnes, water-spatial and FMM.

### III.    EXPERIMENTAL RESULTS

In this section, the results of simulation of the benchmarks execution on the multi-core processor are presented. In the first step, the benchmark execution on a single core processor is simulated. The core utilization and the processor throughput have been captured and shown on two graphs. Then, we simulate the execution of the benchmarks on a four-core processor and put into the graph the utilization and the throughput. In the second step, these graphs are reproduced for each core when benchmarks are run on the multi-core processor. The profiling interval kept fixed or variable in both experiments. It is easy to show the results when the interval is fixed whereas a dynamic interval will reduce the overhead.

### A.  Fixed Interval Length

For each benchmark, three figures are depicted. The first figure shows the results for the benchmark which is run on a processor with a single A-type core. Next two figures are for the first A-type and the first B-type cores in multicore processor respectively.

Figs. 3 to 5 present the simulation results for FMM benchmark. In Fig. 3 that belongs to a single A-type core processor, the throughput-utilization scatter shows a wide distribution of intervals. It is evident from this figure that the intervals could be divided in two or three phases. The throughput scatter indicates repeating in every 100 intervals. Fig. 4 is the similar results for a single B-type processor.

Comparing between Fig. 3 and 4, one can deduce that for the most of intervals the B-type processor has around 10% higher utilization, especially for intervals with throughput more than 600K instruction, and they could belong to the same phase.

Fig. 5 presents the behavior of the FMM benchmark on the first A-type core when it is run on a multicore processor. The results indicate that intervals with high utilization and throughput, could be executed on A-type cores and other intervals could be run on the B-type cores, and this outcome is in consistent with the considerations on figures 3 and 4 for single cores.

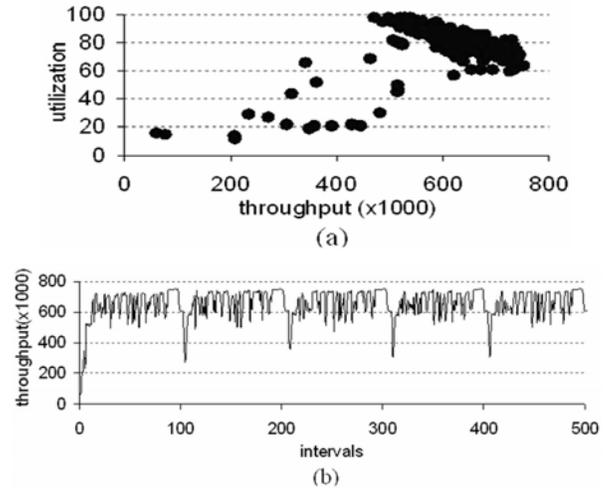

Figure 3.   a)The throughput-utilization and b) throughput graphs for FMM benchmark with 500K cycle interval length on single A-type core processor.

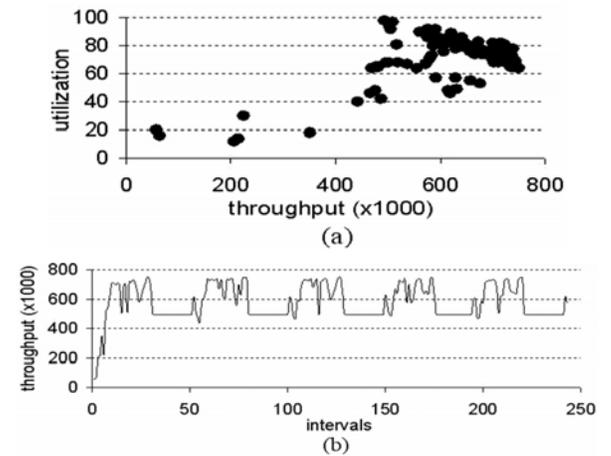

Figure 4.   a) The throughput-utilization and b) throughput graphs for FMM benchmark with 500K cycle interval length on single B-type core processor.

The next considered benchmark is FFT (Figs. 6 to 8). As can be found in the presented results, in comparison with the other investigated benchmark, the distribution of this benchmark is dispersed.

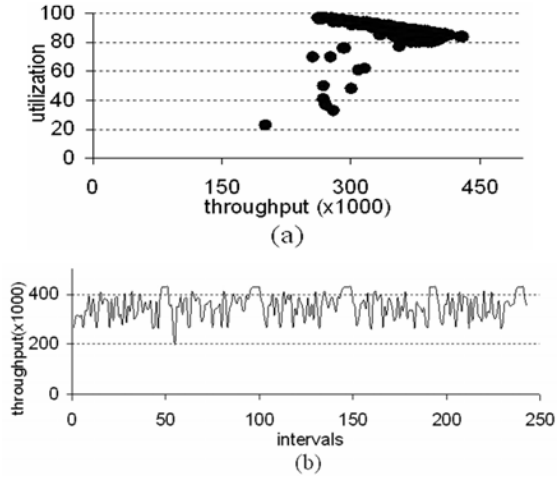

Figure 5.  a) The throughput-utilization b) throughput graphs for FMM benchmark with 500K cycle interval length on the first B-type core of multi-core processor.

The main difference that is clear from the scatter plots is that the FFT benchmark has lower throughput with three distinct phases for both type of processors and cores on multi-core processor. The throughput of the first 30 intervals are for integer instructions that consist the first phase of process. The second phase containing intervals 30 to 185 is composed of integer and floating point instructions. The third phase, from interval number 186 to 270, has both type of instructions, but its integer units' utilization is quite higher than previous phase.

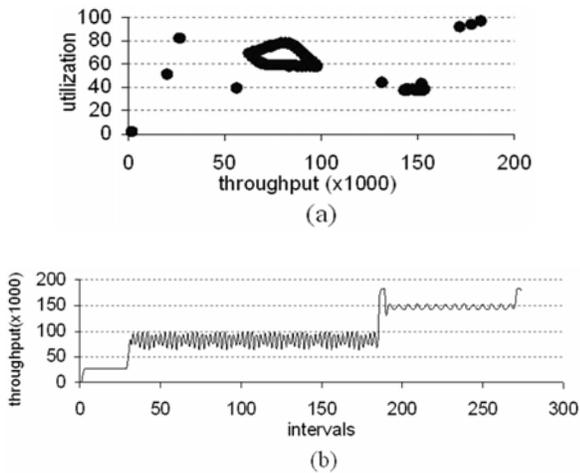

Figure 6.  a) The throughput-utilization and b) throughput graphs for FFT benchmark with 100K cycle interval length on single A-type core processor.

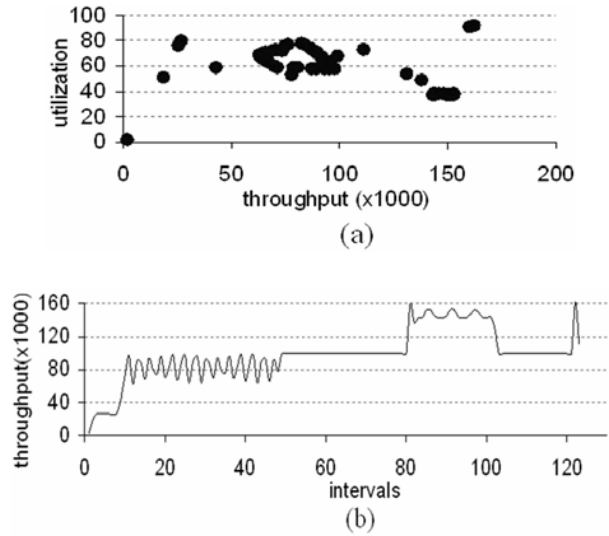

Figure 7.  a) The throughput-utilization and b) throughput graphs for FFT benchmark with 100K cycle interval length on the first A-type core of multi-core processor.

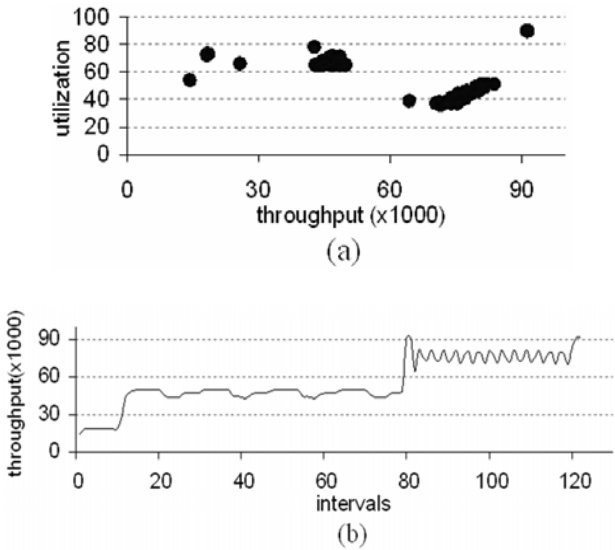

Figure 8.  a) The throughput-utilization and b) throughput graphs for FFT benchmark with 100K cycle interval length on the first B-type core of multi-core processor.

The results for other benchmarks are removed to meet the page limitation of the paper.

### B.  Variable Interval Length

In this section the simulation results for variable interval length are discussed. The throughput scatters of cores in multicore processor are represented. Here a phase change is assumed to accrue when the difference of the throughput of current interval exceeds 100% of the average throughput of all previous intervals (are shown by circles). If the utilization of five successor intervals are more than 95% or less than 30% then a phase change is accrued (shown by triangles). If the difference between the throughput average of the previous intervals and the throughput average of all intervals including the throughput of the current interval stays between -1 and 1 for more than 75 intervals, which shows no change in the application behavior, then the interval length is multiplied by two (shown by squares), otherwise if it exceeds from this range the interval length is divided by two, because it means that application is more likely to change its behavior and needs to be monitored in shorter intervals. The obtained results are depicted in Figs 9 and 10. Note that when the interval length is multiplied by 2, the throughput of the next interval increases as well.

Comparing the results of variable length and the results obtained by applying fix interval length, one can conclude that utilizing the here proposed variable interval length can reduce the profiling overhead more than three times, on average and for all the benchmarks under consideration.

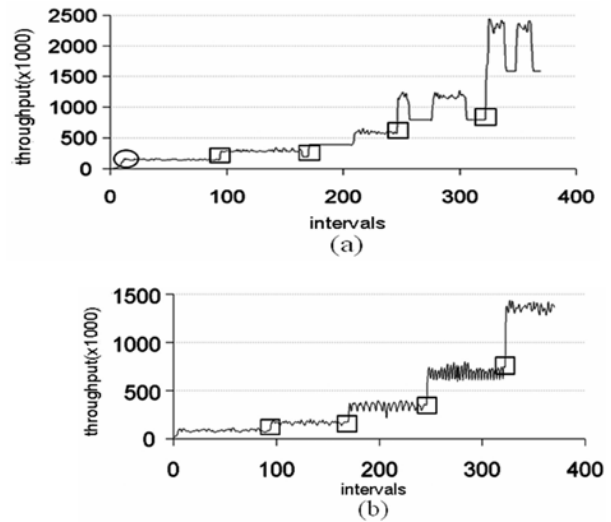

Figure 10. The throughput scatters of variable interval size for a) the firs A-type core and b) the first B-type core for FMM benchmark. The phase changes that are caused by throughput changes and utilization changes are shown by circles and triangles, respectively. Interval length changes are shown by squares.

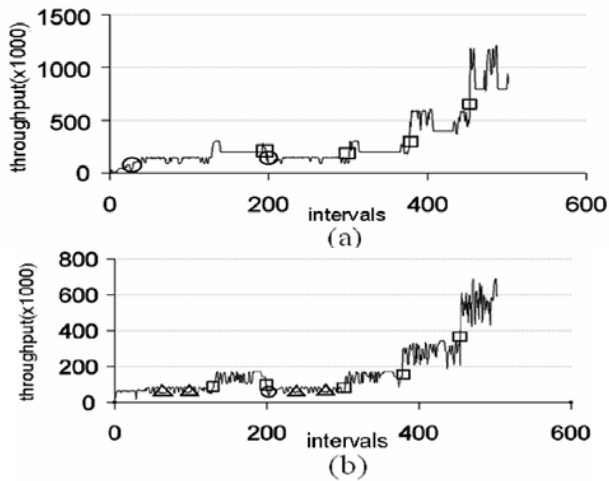

Figure 9. The throughput scatters of variable interval size for a) the firs A-type core and b) the first B-type core for FMM benchmark. The phase changes that are caused by throughput changes and utilization changes are shown by circles and triangles, respectively. Interval length changes are shown by squares.

## IV. CONCLUSIONS

Heterogeneous multicore processors have cores with different level of performances. We can achieve higher throughput and lower power consumption by assigning the incoming process to appropriate core. Moreover, a process may have different resource demands in its different parts. The detection of demand changes can be used in an assignment schedule for the purpose of higher throughput and lower power consumption. In this work we present a theory for detecting demand changes in a process. The detection has been done in dynamic time intervals. The dynamic interval reduces the profiling overheads in the magnitude of three folds.

## AUTHORS PROFILE

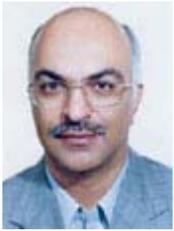

**Morteza Analoui** received his B.S. in Electrical Engineering from Iran University of Science and Technology in 1982. He received his Ph.D. in Electrical and Electronic Engineering from Okayama University in 1990. He is an assistant professor in Iran University of Science and Technology. His research interests include computer networks and Internet, stochastic pattern recognition, grid computing and computer network performance analysis.

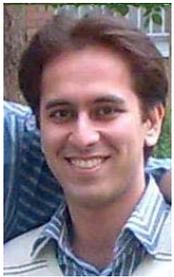

**Ali Z. Jooya** recived his B.S. in Computer Engineering from Azad University in Tehran in 2003. He received his M.Sc. in Computer Engineering from Iran University of Science and Technology in 2009. He worked on dynamic scheduling in heterogenous multicore processors under the supervision of Prof. M. Analoui. His research interest include high performance, low power processor design and moder processor performance evaluation.